
\input phyzzx

\def\rho{\varrho}

\tolerance=500000
\overfullrule=0pt

\pubnum={CPTH/C209-1292}
\date={December 1992}
\pubtype={}
\titlepage

\title{ON THE BREAKDOWN OF PERTURBATION THEORY\footnote{\dag}{Talk at the
International Conference on {\it Modern Problems in Quantum Field Theory,
Strings and Quantum Gravity}, June 1992, Kiev.}}
\vskip 1.0cm
\author{ C.Bachas\footnote{\ddag}{email: BACHAS at ORPHEE.POLYTECHNIQUE.FR}
 }
\address{ Centre de
Physique Theorique \break Ecole Polytechnique \break 91128 Palaiseau,
France.}
\abstract{  The production of $o({1\over g^2})$ particles in a
weakly-coupled
theory is believed to be non-perturbatively suppressed. I comment on the
prospects of {\it (a)} establishing this rigorously, and {\it (b)}
estimating the
effect to exponential accuracy semiclassically, by discussing two
closely-related
problems: the large-order behaviour of few-point Green functions, and
induced
excitation in quantum mechanics. Induced tunneling in the latter case is
exponentially  enhanced for frequencies of the order of the barrier
height.} \endpage

\pagenumber=1

1. {\it Introduction}.
The issue of baryon- and lepton-number violation in the standard
electroweak
model
\REF\tHooft{G. 't Hooft, Phys.Rev. {\bf D14} (1976) 3432.}
\REF\GEN{A. Ringwald, Nucl.Phys. {\bf B330} (1990) 1; \hfil\break
O. Espinosa,  Nucl.Phys. {\bf B334} (1990) 310; \hfil\break
L. McLerran, A. Vainshtein and M.B. Voloshin, Phys.Rev. {\bf D42} (1990)
171.}
[\tHooft] [\GEN]
\REF\Mattis{For a review see M.Mattis, Phys.Rep. {\bf 214} (1992) 159.}
[\Mattis]
has brought back to the limelight the limitations of perturbation
theory.  The problem, in a nutshell, is that naive
 weak-coupling expansions cannot be used
to estimate processes in which a large number of particles
 is involved.
If we denote for instance  by $N\equiv {\nu\over g^2}$
 the number of $W$ bosons produced in
 a high-energy collision, then an expansion of the inclusive
cross-section in terms of Feynman diagrams
 corresponds to expanding simultaneously
in the gauge coupling $g$ and in $\nu$.
 This is of no use if one is interested
in the region $g\rightarrow 0$ with $\nu$ held fixed and not small.
Indeed the leading-order result for this process, whether accompanied
[\GEN] or not
\REF\GOLD{J.M. Cornwall, Phys.Lett. {\bf 243B} (1990) 27;\hfil\break
H. Goldberg, Phys.Lett. {\bf 246B} (1990) 445.} [\GOLD]
by vacuum tunneling, violates for large $\nu$ the unitarity
bounds, and is hence manifestly unreliable.
A breakdown of perturbation theory of course also occurs in
 other contexts, such
as at high temperature, high densities or large external
fields. In all these cases it is however obvious that the
 properties of the
perturbative vacuum are being   modified drastically,
 so that the theory is no more weakly-coupled. In high-energy
two-particle collisions on the other hand, our
difficulty in calculating multiparticle production
 looks more like a technical
nuisance, rather than a signal that these processes are unsuppressed.
This sounds intuitively obvious if one thinks of the time-reversed process:

 how
could we send $50$ $W$   bosons in an interaction region, and expect only
an
energetic $e^+$$e^-$ pair to emerge?\hfil\break
\indent There has been much learned debate on this issue, which I could not
possibly review in this short talk.
Here I will focus briefly on
two closely-related problems where definitive
answers are available: the problem of induced excitation in quantum
mechanics,
and the large-order behaviour of typical Euclidean Green functions.
Based on these I will then make
  a few comments
on  the prospects
 {\it (i)} of proving that   multiparticle
production in a weakly-coupled theory is  non-perturbatively small, and
{\it (ii)} of  actually calculating it to exponential accuracy in a
semiclassical approximation.
Given the  intuitive
evidence for exponential suppression,
and the ridiculously small low-energy tunneling
probability   in the electroweak model,
$e^{-{4\pi\over \alpha_w}} \sim 10^{-156}$,
 one may   wonder whether calculating zero to exponential accuracy
is really worth all this effort.
I think the answer is yes for two reasons:
  first, the issue is sufficiently important
  to   deserve  that we close
all loopholes in our intuitive arguments.
Second,  learning how to
calculate  such non-perturbative phenomena
may turn out to be academic in the electroweak
model, but very interesting in other contexts.
In this respect simpler models could be more valuable
for their own sake, rather than as paradigms
 of multi-$W$ and -$Z$ production.
\vskip 2cm

2. {\it Large-order behaviour}.
We are all well-accustomed to the fact that in field theory perturbative
expansions are usually divergent. Indeed, a finite radius of convergence
would
be in contradiction with the vacuum instability that typically develops for
negative square coupling
\REF\Dyson{F.J.Dyson, Phys.Rev. {\bf 85} (1951)
631.}[\Dyson]. Consider for definiteness a scalar field in $d$ dimensions,

whose action in appropriate mass units is
$$S =  \int d^dx\ \Bigl[{1\over 2}(\partial\phi)^2 +
{1\over 2} \phi^2 + {g^2\over 4}
 \phi^4\Bigr]  \   , \eqno(1)$$
and
let
$$ G^{(N)}(g^2 \vert \  p_1, ..., p_{N})  \asymp \sum_{n=0}^\infty
 g^{2n} G^{(N)}_n(p_1, ..., p_{N})
  \eqno(2)$$
be the expansion of a Euclidean $N$-point   function.
For $n\gg 1$ we can compute the coefficients of the series semiclassically
  \REF\Lipatov{C.M. Bender and T.T. Wu, Phys.Rev.{\bf
D7} (1973) 1620; \hfil\break
L.N.Lipatov, Sov. Phys. JETP {\bf 44} (1976) 1055, and {\bf 45}
(1977) 361;\hfil\break
E. Br\'ezin, J.C. Le Guillou and J. Zinn-Justin, Phys.Rev.{\bf
D15} (1977) 1544, 1558;\hfil\break
G. Parisi, Phys.Lett. {\bf 66B} (1977) 167.}
  [\Lipatov]
\REF\Zinn{J.C. Le Guillou and J. Zinn-Justin editors,
{\it Large-Order Behaviour
of Perturbation Theory}, North Holland (Amsterdam 1989);\hfil\break
J. Zinn-Justin, {\it Quantum Field Theory and Critical Phenomena}, Oxford
U.
Press 1989. } [\Zinn].
The idea is to relate them  by a dispersion integral  to the discontinuity
of
the function on the negative-$g^2$ axis,
$$  G^{(N)}_n(p_1, ..., p_{N}) = {1\over 2\pi i} \int_{-\infty}^{0} {dg^2
\over
g^{2n+2}}\ {\rm disc}\  G^{(N)}(g^2 \vert\  p_1, ..., p_{N}) \eqno(3)$$
The integral is then evaluated at the saddle point
$ \phi_{cl}(x)\equiv{1\over \sqrt{-g^2}} f(x)$, which describes the decay
of
the vacuum for negative square coupling, with the result
$$ G^{(N)}_n(p_1, ..., p_{N}) \ \sim_{_{\!\!\!\!\!\!\!\!\!
n\rightarrow\infty}}
\ (2\pi)^{{d\over 2}-1} \Gamma\Bigl(n+{N+d\over 2}\Bigr)\ (-)^n
a^{-n-{N\over 2} }
\prod_{j=1}^N f(p_j)\   \ \ .\eqno(4)$$
Here   $S(\phi_{cl})\equiv{a\over -g^2}$ is the saddle-point
action, and we have  dropped a
  momentum-conserving $\delta$-function.
Note that a   weak-coupling expansion of the discontinuity
  is justified in the large-$n$ limit, provided $N$ and $p_j$
are all kept finite and fixed.

The rapid factorial growth  of the coefficients,
typical of bosonic field theory,
makes the series (2) diverge for any $g$.
A closer look at Feynman diagrams
in fact reveals
that these large contributions at high
orders arise when $n\sim o({1\over g^2})$ vertices are
 concentrated in a real-space
region of size
 $\sim o(1)$ \footnote*{Fermionic theories
 fare in this respect better thanks to the Pauli
exclusion principle. In closed-string theory, on  the other hand,
the divergence of perturbative expansions appears
to be even more severe
\REF\Gross{D. Gross and V. Periwal, Phys.Rev.Lett. {\bf 60} (1988) 2105;
\hfil\break
 S. Shenker, in {\it Random Surfaces and Quantum Gravity}, O. Alvarez et al
eds., Plenul Press, New York 1991. } [\Gross].}.
{}From the functional-integral point of view these contributions probe
  the region of {\it large} fields, or of large virtual {\it non-linear}
 waves. It is because such waves are not
    described adequately,
when one expands around the usual vacuum, that perturbation theory breaks
down.
 Intuitively we of course expect that large fluctuations,
 though perhaps hard to
calculate precisely, are nevertheless {\it
exponentially suppressed} and thus often negligible. This expectation is in
practice confirmed
 by the enormous success of QED:
keeping for instance three terms in the expansion of the
anomalous magnetic moment of the electron,
we find agreement with experiment to
better than seven significant digits
\REF\Kino{T. Kinoshita and W.B. Lindquist, Phys.Rev.Lett. {\bf 47} (1981)
1573.}
[\Kino].
The more precise mathematical statement one would like to make is that the
  perturbative series is   asymptotic  to some rigorously defined
function,  and that the semiclassical estimate, eq.(4), can be converted
into a {\it uniform  bound}   for the series remainder:
$${\rm Rem}_n (G) \equiv \Bigl\vert
G^{(N)}(g^2) - \sum_{l=0}^{n-1} G^{(N)}_l g^{2l}\Bigr\vert\ < \  n!\
   g^{2n} \ a_n
  \   , \eqno(5)$$
  where    $\lim_{n\rightarrow\infty}  (a_n)^{-1/n} = {\tilde a}$  is a
finite constant, and $g^2$ lies in some
finite interval $(0,{\tilde g}^2)$ on the positive real axis.
  From the above   bound and from
Stirling's
formula, $n! \simeq \sqrt{2\pi}(n+1)^{n+{1\over
2}}e^{-n-1}$, we  could conclude that in the $g\rightarrow 0$ limit
 $$ \min_n\  {\rm Rem}_n (G) \ \preceq
 {\rm exp}\Bigl({-{{\tilde a}\over g^2}}\Bigr)  \ \ ,\eqno(6)$$
so that the divergent high-order contributions    indeed sum up into
a controllable non-perturbatively small ambiguity
\footnote*{The simple statement of asymptoticity of the series
says nothing about the size of this ambiguity,
 which could for instance have been as
large as ${\rm exp}(-1/g^{1\over 100})$.}.

Proving this statement
has been one of the aims of formal (constructive) field
theory
\REF\Riv{J. Glimm and A. Jaffe, {\it Quantum Physics, a functional integral
point of view}, McGraw-Hill, New York 1981; \hfil\break
  V. Rivasseau, {\it From Perturbative to Constructive Renormalization},
  Princeton
U. Press, Princeton 1991.}
[\Riv]. For the Euclidean scalar theory, eq.(1), in the superrenormalizable
($d<4$) domain, the stronger result
of Borel summability
\REF\Borel{ S. Graffi, V. Grecchi and B. Simon, Phys.Lett. {\bf 32B}
(1970); \hfil\break
J.P. Eckmann, J. Magnen and R. S\'en\'eor, Comm.Math.Phys. {\bf 39} (1975)
251;
\hfil\break
J. Magnen and R. S\'en\'eor, Comm.Math.Phys. {\bf 56} (1977) 237.}
[\Borel] ensures in particular the existence of the above bound
\footnote{\ddag}{Borel summability in fact allows a reconstruction of the
function
from its series with arbitrary precision,
by using appropriate conformal mappings
\REF\Borell{B. Hirsbrunner, Helv.Phys.Acta
{\bf 55} (1982) 295.}[\Borell] [\Zinn]
.}.
Furthermore,
  the large-order semiclassical estimate can be shown to correctly predict
  the relevant singularity of the Borel transform
  \REF\Rivv{J. Magnen and V. Rivasseau, Comm.Math.Phys. {\bf 102} (1985)
59;\hfil\break
J. Feldman and V. Rivasseau, Ann. Inst. H. Poincar\'e, {\bf 44} (1986)
427.}
[\Rivv], so that the bound is optimized for
  ${\tilde a} = a$.
The extension of these results to {\it (i)} the Minkowski region in field
theory,
{\it (ii)} the double-well potential in quantum mechanics, or
{\it (iii)} the $d=4$ renormalizable theory
 is not
straightforward. Controlling non-perturbative effects  in the first case
 is hard because
  bounds do not continue analytically.
Borel summability has nevertheless been
established for the
on-shell four-point function,
 but only below the particle-production threshold
\REF\Epstein{J.-P. Eckmann and H. Epstein,
Comm.Math.Phys. {\bf 68} (1979) 245.}
[\Epstein].
In the case of the
  double-well potential
of quantum mechanics, obtained by flipping the sign of the quadratic
term in (1), there is a hindrance to Borel summability
  due  to the presence of real instantons.
Indeed the semiclassical large-order analysis
\REF\Br{E. Br\' ezin, G. Parisi and J. Zinn-Justin, Phys.Rev. {\bf D16}
(1977)
408.}[\Br]  predicts a singularity
of the Borel transform at the same distance, $a={4\over 3}$, from the
origin as in the single-well case, but lying on the positive real axis.
This singularity corresponds to the action of an infinitely-separated
instanton-antiinstanton pair, and is expected to control the
non-perturbative
ambiguities of the asymptotic expansions in the trivial and one-instanton
sectors [\Zinn], but   rigorous bounds of the type (5) have not to my
knowledge
been established \REF\Simon{ B.Simon,
Int.J.Quant.Chem. {\bf XXI} (1982) 3;\hfil\break
E. Caliceti, V. Grecchi and M. Maioli, Comm.Math.Phys. {\bf 113}
(1988) 625.}[\Simon].  In the third case of the  $\phi^4_4$  model, Borel
summability is obstracted by the so-called
  renormalons
\REF\ren{G. Parisi, Phys. Rep. {\bf 49} (1979) 215; \hfil\break
G. 't Hooft, in {\it The whys of subnuclear physics}, ed. by A. Zichichi,
Plenum
Press, New York 1979.
}[\ren], which are a warning that non-perturbative ambiguities could a
priori
be affected by renormalization. Since, however, the theory most probably
does not exist, there is little point in
worrying about   bounds (5).
Needless to say, finally, that non-perturbative effects
in four-dimensional gauge theories are still beyond rigorous technical
control [\Riv], since in addition to renormalization and the presence of
real
instantons one must also  face the problems of gauge fixing and scale
invariance.

There are three lessons to retain from this blitz review of large-order
behaviour:
first that naive perturbation theory around the vacuum breaks down when one
tries
to estimate the contribution of large fields, or of (real or virtual)
non-linear
waves. Second that such contributions to few-particle processes are
expected to
stay exponentially small, and can be estimated to exponential accuracy
semi-classically.
 And third that a real proof of exponential suppression requires a
non-perturbative control of the vacuum and takes us into the rough
territory of
constructive field theory. These comments should be kept in
mind when moving on to
the problem of multi-particle production, which in a vague sense is the
{\it
square root} of the large-order problem: indeed, large-field
 fluctuations are in this case created from, but do not have to disappear
 back to a
few-particle state.

 \vskip 0.8cm

3. {\it Multi-leg functions in Quantum Mechanics}.
    The quantum mechanical analog of multiparticle production is the
induced
excitation of an anharmonic oscillator  under the action
of a weak but very energetic external force
\REF\WKB{M.B. Voloshin, Phys.Rev {\bf D43} (1991)
1726;\hfil\break M. Porrati and L. Girardello, Phys.Lett. {\bf B271} (1991)
148;
\hfil\break
J.M. Cornwall and G. Tiktopoulos,  Phys.Lett. {\bf B282} (1992) 195.}
 \REF\BLST{C.Bachas, G.Lazarides, Q.Shafi and G.Tiktopoulos,
Phys.Lett.{\bf 268B} (1991) 401;\hfil\break
V.G.Kiselev, Phys.Lett. {\bf B278} (1992) 454\hfil\break
G. Diamandis, B. Georgalas, A. Lahanas and E. Papantonopoulos, UA/NPPS-8-92
(August 1992) .} \REF\Bac{C. Bachas, Nucl.
Phys. {\bf B377} (1992) 622. }[\WKB] [\BLST] [\Bac].
We concentrate first on the single-well potential with action given by
eq.(1).
The   amplitude of interest is
proportional to the matrix element
$\langle 0\vert \phi \vert N\rangle$,
where  $N \equiv {\nu\over g^2}\gg 1$  is the
level of the excited final state and
$E  \equiv {\epsilon\over g^2}$ is its energy.
To motivate this scaling of parameters consider the
 weak-coupling expansion of the
energy,
$$\eqalign{ E  =  (N + {1\over 2}) + {3  g^2\over 16} (2 N^2 &+ 2 N + 1 )
\cr  - &{g^4\over 128} ( 34 N^3 + 51 N^2 + 59 N + 21 ) + ...
\cr}\eqno(7a)$$
which can be clearly  reorganized as follows:
$$ \epsilon =  \epsilon(\nu) + g^2 \epsilon_1(\nu) + g^4 \epsilon_2(\nu) +
...
\eqno(7b)$$
The effective expansion parameter in $(7b)$ is Planck's constant, as would
be
made obvious if we were to restore all units in our equations.
In particular, the
function $\epsilon(\nu) = \nu + {3\over 8}\nu^2 + {34\over 128}\nu^3 + ...$
is
the inverse of the integrated   classical  density of states,
$$ \nu(\epsilon) = {2\over \pi} \int_0^{x(\epsilon)} dx\
\sqrt{2\Bigl(\epsilon -  {x^2\over 2} - { x^4\over 4}\Bigr)} \eqno(8)$$
where $\phi\equiv x/g$ defines the rescaled position variable, and
$x(\epsilon) = \sqrt{\sqrt{1+4\epsilon}-1} $ is the rescaled   classical
turning
point. This relation
between energy and level illustrates
  that, for coherent states of many quanta,  it is the
{\it classical}  but not the naive {\it weak-coupling} approximation that
is
adequate.

Let us then consider an analogous semiclassical expansion of the
matrix-element  $\langle N'\vert x^M \vert N\rangle$ , when all quantum
numbers
$N \equiv {\nu\over g^2} $,
$N' \equiv {  \nu'\over g^2} $,
and $M \equiv {\mu\over g^2} $    are
    large \footnote{\dag}{Because of reflection symmetry  $N+N'+M$ must be
even.},
 $$\langle N'\vert x^M \vert N\rangle = {\rm exp}\Bigl( \ {1\over g^2}
 F(\nu,\nu',\mu) + F_1(\nu,\nu',\mu) +
  ...\Bigr) \ \ .\eqno(9)$$
The  form of this expansion is based on
our expectation that in some range of parameters
the overlap integral should be
non-perturbatively suppressed. We are of course ultimately interested in
the
matrix element $\langle 0\vert \phi \vert N\rangle$, since we want to mimic
the effect of a few-particle initial state. We must therefore hope that
this matrix element can be obtained, at least to exponential accuracy, by
taking
the $ \mu,\nu'\rightarrow 0$ limit, i.e. in formulae:
$$ \lim_{g\rightarrow 0}\  g^2 {\rm log}\langle 0\vert \phi \vert N\rangle
=
F(\nu)\ \ ,\eqno10)$$
where
   $$ F(\nu) \equiv \lim_{\mu, \nu'\rightarrow 0} F(\nu,
\nu', \mu)\ \ .\eqno(11)$$
The simple unitarity relation
$$ \langle 0\vert \phi^2 \vert 0\rangle = \sum_{m=0}^{\infty} \ \vert
\langle
0\vert \phi \vert m\rangle \vert^2
 \eqno(12)$$
implies in particular that
 $\vert \langle 0 \vert \phi\vert N\rangle \vert^2 <  \langle 0\vert \phi^2
\vert
0\rangle \simeq {1\over 2} + o(g^2)$ , so that the function $F(\nu)$ must
clearly
stay {\it non-positive}. The first term of a
naive weak-coupling expansion violates
this unitarity bound,  because of the same factorial growth of graphs
which is responsible for the
large-order divergence. A precise calculation in fact
gives
 \REF\Tree{ M.B. Voloshin, TPI-MINN-92/38-T (August 1992); \hfil\break
E.N. Argyres, R. Kleiss and C. Papadopoulos, CERN-TH. 6496  (May 1992).}
 [\Tree]
\footnote*{The calculation is identical to the Born approximation for the
production of $N$ scalar particles at threshold, up to a normalization
${1\over
\sqrt{N!}}$, and an extra factor of
$\sqrt{2}$ per external leg  appearing in the LSZ
reduction of quantum mechanics
[\Bac].} $$ \langle 0 \vert \phi\vert N\rangle  = \sqrt{8\over g^2}\
\sqrt{N!}
\Bigl({g^2 \over 16}\Bigr)^{N/2} \ [1 + o(g^2)] \ , \eqno(13a)$$
from which we find
$$ F(\nu) = -{\nu\over 2} (1+log{16\over \nu}) +
o(\nu^2 log\nu) \ \ .\eqno(13b)$$
The Born approximation  thus hits the unitarity bound at
$\nu=16e$, at which point it is manifestly unreliable. In reality, however,
the
matrix element stays exponentially small for all finite $\nu$. This can be
  established explicitly both {\it (a)} by a semiclassical calculation
\REF\Landau{L.D. Landau and E.M. Lifshitz, {\it Quantum Mechanics,
Non-Relativistic Theory}, 3rd edition, Pergamon Press}
\REF\Ti{J.M. Cornwall and G. Tiktopoulos, in
preparation.}
[\Landau] [\WKB] [\Ti], and {\it (b)} by deriving rigorous
 upper  bounds [\Bac], as I will now briefly explain:

{\it (a)}
  The semi-classical calculation of the
overlap integral $\int  \  x^M \Psi_{N'}^{\dag} \Psi_{N }$ for $N\gg N'\gg
1$
was suggested
a long time ago by Landau
[\Landau]. The idea is
to deform the integration contour
in the complex-$x$ plane
away from the classical
turning points, use a WKB approximation for the wavefunctions $\Psi_{N }$
and
$\Psi_{N'}$, and evaluate
the resulting integral at its saddle point. The full details of a carefull
calculation have only recently been completed [\Ti], and confirm Landau's
result
[\WKB, \Landau] in the $\mu\rightarrow 0$ limit:
$$\eqalign{ F(\epsilon, \epsilon') =& - \int_{\epsilon'}^{\epsilon } du
\int_{x(u)}^{\infty} {dx \over \sqrt{2 ({x^2\over 2} + { x^4\over 4}-u)}}
\cr
=&  - \int_{\epsilon'}^{\epsilon } {du \over (1+4u)^{1\over 4}} {\bf K}\Bigl(
\sqrt{1+\sqrt{1+4u} \over 2\sqrt{1+4u}}\Bigr) \cr } \eqno(14)$$
where
${\bf K}$ is   the complete elliptic integral.
The result is here given as a function of   energies,
but can be also expressed in
terms of the levels $\nu$ and $\nu'$ via the classical relation (8).
As was already anticipated, $F$
has a smooth $\nu'\rightarrow 0$ limit,
which can be used to estimate $\langle 0 \vert \phi\vert N \rangle$ to
exponential accuracy.
Furthermore $F(\nu)$ is a monotone decreasing function
of energy or level, so that the matrix element for finite $\nu$ is
indeed exponentially small.
  Notice also that by using the
asymptotic expansion
${\bf K}(1-\delta) \sim log(\sqrt{8\over \delta}) + o(\delta\ log\delta)$,
one
can recover the Born result, eq.$(13b)$.

{\it (b)} The more rigorous proof of exponential suppression [\Bac] makes
use
of an exact recursion relation between matrix elements of powers of $\phi$.
Defining the recursion coefficients $ \langle N' \vert
\phi^M \vert N \rangle \equiv
{\cal R}_M^{(N,N')} \  \langle N'\vert \phi^{M+2} \vert N \rangle  $
one finds
\footnote{\dag}{This equation allows
a fast numerical evaluation of the low-lying energy
levels and wavefunctions with very high precision
\REF\Blank{J.L. Richardson and R. Blankenbecler, Phys.Rev. {\bf D19} (1979)
245.}
[\Blank].
Notice also that in the semiclassical limit it reduces to
$${\mu^4\over 6}\eta^3 + \mu^2 (\epsilon+\epsilon')\ \eta^2 +
[(\epsilon-\epsilon')^2 - \mu^2]\ \eta - {\mu^2\over 2} = 0 $$
where
$-{1\over 2}log\eta \equiv {\partial F \over
\partial\mu}(\epsilon,\epsilon',\mu)$. It can therefore
be used to determine the
$\mu$-dependence of the function $F$ completely.}:
$$ {\cal R}_M \ = \ {g^2 {M+1
\choose 2} \over 6{M\choose 4}{\cal R}_{M-2}
{\cal R}_{M-4} + 2{M\choose 2}(
E + E'){\cal R}_{M-2} + ((E-E')^2-M^2)}\ , \eqno(15) $$
where  ${a\choose b}$ are the binomial coefficients, and
$E$ and $E'$ are the energies at the $N$ and $N'$ levels.
This is the analog of Schwinger-Dyson equations, which in field theory
are instrumental for proving the uniform remainder bounds (5).
The problem we are here facing is very similar: indeed,
since the first $N$ terms
in the asymptotic power series for $\langle 0 \vert \phi\vert N\rangle $
are zero, we have for all $n<N$
$$ {\rm Rem}_n \bigl( \langle 0 \vert \phi\vert N\rangle \bigr) =
\langle 0 \vert \phi\vert N\rangle \ ,\eqno(16)$$
so that
bounding
the series remainder up to this order would automatically
put bounds on the matrix
element itself. Skipping further details on the derivation of these
bounds, let me simply point out that they finally take the form
$$ F(\epsilon) \ \le \ \min_{0\le\xi\le{\epsilon\over 2}} \ B(\epsilon,\xi)
\ ,
\eqno(17)$$
where $ \xi/g^2$ plays the role of the order of the expansion. This should
be
chosen appropriately so as to optimize the bound $B$, making sure in
particular
that the coupling-constant suppression is not overwelmed by factorial
growth.
The optimal bound has been shown to decrease monotonically
with energy [\Bac],  like the
semiclassical estimate eq.$(14)$. Proving that the two actually coincide
would
require some more work.

\vskip 0.5cm

The double-well potential,
  $V(\phi) =  = \  {g^2 \over 16}\
\bigl (\phi^2-{2\over g^2}\bigr )^2 $, can be treated with similar
techniques.
Induced excitation, whether accompanied or not by
quantum tunneling,
can be again bounded by an exponential envelope whose decay with energy is
monotonic [\Bac]. The leading-order instanton calculation [\BLST], on the
other hand,
predicts
that induced tunneling grows fast with energy in the low-energy region.
To be more precise, the
transition amplitude from the ground
state in one well to the $N{\rm th}$ excited state in the other is given
to exponential accuracy,  and in the limit of small $\nu$, by
$$F(\nu) \ = \ -{4\over 3} + {\nu \over 2}
(1+log{16\over \nu}) + {\rm subleading}\ \ .
\eqno(18)$$
This rapid initial growth with energy is the result of two competing
effects: the
enhancement of spontaneous tunneling, starting from some optimal jumping state,
is partially compensated by the fast-falling probability of exciting this
latter
from the vacuum \REF\Banks{T. Banks, G. Farrar, M. Dine, D. Karabali and B.
Sakita,
Nucl. Phys. {\bf B347} (1990) 581.}[\Banks].
The presence of extra real turning points complicates in this case the
WKB analysis, and the calculation of the full function
$F(\nu)$   has not yet been completed [\Ti]. The exponential
envelope tells us, however,   that it has to
reach  a maximum at some finite
distance below the zero-axis, before dropping indefinitely
in the $\nu\rightarrow \infty$ limit.
This means that induced tunneling has a {\it resonance of exponential
proportions}, which could, for instance, be observable
 in semi-conductor physics \REF\semi{See for instance the reviews by
E.Esaki,
IEEE J.Quantum Electron. {\bf 22}(1986) 1611, and by F.Capasso, K.Mohammed
and
A.Y.Cho, ibid {\bf 22}(1986) 1853;\hfil\break
also G.Livescu et al, Phys.Rev.Lett.
{\bf 63} (1989) 438, and refs. therein.} [\semi].

\vskip 0.4cm
4. {\it Comments on Field Theory}.
 I will conclude with a few telegraphic comments
 on the prospects of extending
the results reviewed in the previous
two sections to multi-particle production in field theory.
We saw that the large-order behaviour shows no qualitative difference as
one
passes from ($d=1$) quantum mechanics, to ($d=2,3$) superrenormalizable
Euclidean  theories.
It should therefore be possible   to
prove that, like induced excitation in the former,
multiparticle production in the latter  is exponentially
suppressed.
Repeated use of the Schwinger-Dyson equations
could indeed lead to bounds for Euclidean
multi-leg Green functions, but there
is at present no obvious strategy on how to extend such bounds
to the Minkowski region.
Closely related are efforts to
establish bounds by exploiting the relation
between  the forward elastic amplitude and
the total inclusive cross-section
\REF\Ven{V. Zakharov, Nucl.Phys. {\bf B353} (1991) 683; \hfil\break
G. Veneziano, unpublished note (1990) and Mod.Phys.Lett. {\bf A7} (1992)
1667;\hfil\break
M. Maggiore and M. Shifman, Nucl.Phys.
{\bf B371} (1992) 177. \hfil } [\Ven]
$$ \sigma_{incl}(\sqrt{s}) = {1\over \sqrt{(s-4) s}} {\rm Im} {\cal
A}_{el}(\sqrt{s})
\ .\eqno(19)$$
This is  the analog of the quantum-mechanical
unitarity relation, eq.($12$).
  Taking the remainder at order $N$, we
can express the
inclusive cross section for producing at least
$N$ particles in the following form:
$$ \sigma_{n\ge N}(\sqrt{s}) = {1\over \sqrt{(s-4) s}}
{\rm Im}\  {\rm Rem}_N
\bigl({\cal A}_{el}\bigr) + {\rm Rem}_N
\bigl(\sigma_{n<N} \bigr) \ \ . \eqno(20)$$
Rigorous bounds on the large-order behaviour of the right-hand side could
thus be used to establish exponential suppression of the left-hand side.
Though
very plausible, this strategy stumbles again on the fact that little can be
said
rigorously about the large-order behaviour of perturbation theory deep in
the
Minkowskian region.

In view of these difficulties, attempting to calculate the process
semi-classically
\footnote{\dag}{That a part of this problem, i.e. the calculation of
final-state
corrections, admits a semi-classical expansion has been known for a while
\REF\Yaf{S.Yu. Khlebnikov, V.A. Rubakov and P.G. Tinyakov, Mod.Phys.Lett.
{\bf A5}
(1990) 1983; Nucl.Phys. {\bf B350} (1991) 441; \hfil\break
L. Yaffe, in {\it
Baryon Number Violation at the SSC?}, Proc. of the Santa Fe Workshop,
M.Mattis and E.Mottola eds., World Scientific, 1990 ; \hfil\break
P. Arnold and M. Mattis, Phys.Rev. {\bf D42} (1990) 1738.\hfil}[\Yaf]
[\Mattis].}
remains, I believe, the most promising prospect.
This prospect has received a boost recently thanks to a clever suggestion
by  Rubakov and Tinyakov
\REF\Rub{V.A. Rubakov and P.G. Tinyakov, Phys.Lett. {\bf 279} (1992)
165;\hfil\break
 P.G. Tinyakov, Phys.Lett. {\bf 284} (1992) 410.\hfil}
 [\Rub], who proposed to first
  calculate a quantity
 which appears to have a manifest semi-classical expansion.
 This quantity is the inclusive cross-section for an ensemble of
  $N'\equiv{\nu'\over g^2}$   incoming particles, distributed
randomly in phase space, but with fixed total energy in the center-of-mass
frame.
Assuming we can calculate it, we may then hope to take the
$\nu'\rightarrow 0$ limit, so as to recover the leading exponential
estimate for processes
with only
few particles in the initial state.
This has been illustrated explicitly in our
discussion of the single-well potential in
quantum mechanics. There is, furthermore, some preliminary evidence for the
smoothness of this limit in the one-instanton sector of the
standard electroweak model
\REF\Mue{A. Mueller, CU-TP-572 preprint (August 1992).}[\Mue].
Whether a useful solution to this complicated saddle-point problem
  can, however,
be found remains   to be seen.

I have benefited from discussions with  K. Gawedzki, G. Grunberg, J.
Lascoux,
A. Mueller, E. Mottola, P. Tinyakov
E. Papantonopoulos and
R. Seneor. I am particularly
indebted to V. Rivasseau and J. Magnen for many patient explanations of
  constructive field theory methods.

\nobreak

\refout
\vfil
\end